\documentclass[a4paper]{jpconf}
\usepackage{a4wide}
\usepackage{graphics}
\usepackage{graphicx}
\usepackage{epsfig}
\usepackage{amssymb}
\usepackage{amsthm}
\usepackage{lineno}
\usepackage{amsmath}
\usepackage{color}

\newcommand{\Xp}{\Xi_{cc}^+}
\newcommand{\Xpp}{\Xi_{cc}^{++}}
\newcommand{\Lc}{\Lambda_c^+}
\newcommand{\Jp}{J/\psi}
\newcommand{\MeV}{\mbox{\,MeV}}
\newcommand{\GeV}{\mbox{\,GeV}}

\begin{document}
\title{Signals of the double intrinsic heavy quark at the current experiments}
\author{Sergey Koshkarev, Stefan Groote}

\address{Institute of Physics, University of Tartu, 10139, Estonia}

\ead{sergey.koshkarev@ut.ee}

\begin{abstract}
In this paper we review existing data and analyze experimental opportunities
for modern experiments to investigate the signals of intrinsic heavy quarks.
\end{abstract}

\section{Introduction}

Even though the existence of intrinsic heavy quarks was proposed more than 35
years ago, the mechanism is still under construction~\cite{Brodsky:1980pb}.

The existence of heavy quarks in the proton's light-front (LF) wavefunction at
a large LF momentum fraction $x$ is in fact predicted by QCD if one analyzes
the higher Fock states $|uud c\bar c\rangle$ and $|uud c\bar cc\bar c\rangle$
in the hadronic eigenstate, i.e., Fock states where the heavy quark pairs are
multi-connected to the valence quarks. LF wavefunctions, the eigensolutions of
the QCD LF Hamiltonian, are defined at fixed LF time $\tau=t+z/c$ and are thus
off-shell in the invariant mass. In QED for example, positronium has an
analogous $|e^+e^-\mu^+\mu^-\rangle$ Fock state due to the insertion of
light-by-light scattering in the positronium self-energy amplitude.

In such an ``intrinsic charm'' Fock state $|uudc\bar c\rangle$, the maximum
kinematic configuration occurs at minimum invariant mass where all quarks are
at rest in the hadron's rest frame, i.e., at equal rapidity in the moving
hadron. Equal rapidity implies $x_i\propto(m^2+{\vec k_\perp}^2)^{1/2}$ for
each quark, so that the heavy quarks in the Fock state carry most of the
hadron's LF momentum. The operator product expansion predicts that the
probability of intrinsic heavy-quark Fock states $|uud Q\bar Q\rangle$ 
scales as $1/m_Q^2$ due to the non-Abelian couplings of
QCD~\cite{Brodsky:1984nx,Franz:2000ee}.

\section{Review of existing data}

\begin{figure}[t]
\kern-16pt\includegraphics[scale=0.35]{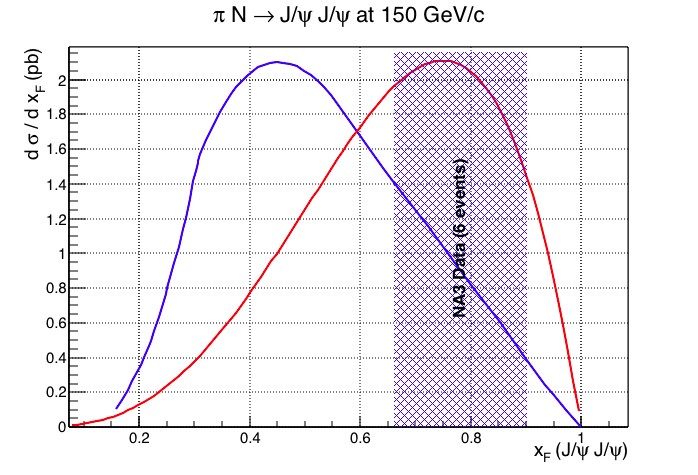}
\kern-16pt\includegraphics[scale=0.35]{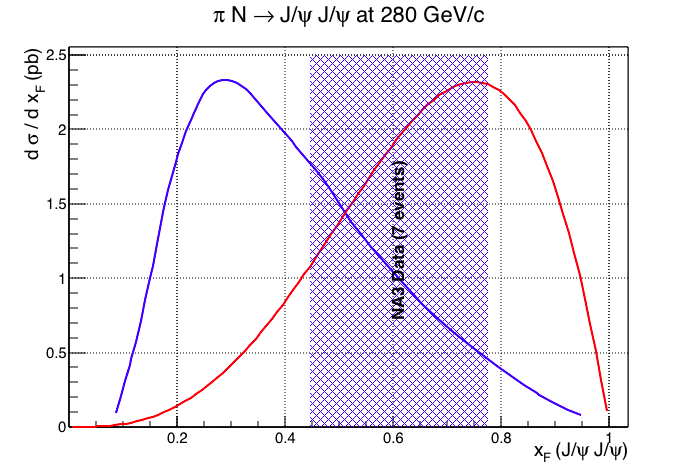}
\caption{\label{fig:na3data} NA3 events (shaded area), pQCD
prediction~\cite{Ecclestone:1982yt} (blue left curve) and prediction of the
heavy quark mechanism (red right curve) at the $150\GeV/c$ (left panel) and
at the $280\GeV/c$ $\pi$ beam (right panel)}
\end{figure}

Using the CERN pion beam at $150$ and $280\GeV/c$ to produce charm particles
with incident on hydrogen and platinum targets, the NA3 experiment provided
data on the production of the double $J/\psi$ with very similar features: a
high value for the ratio
$\sigma(\psi\psi)/\sigma(\psi)=(3\pm 1)\times 10^{-4}$, values
$x_{\psi\psi}>0.6$ at $150\GeV/c$ and $x_{\psi\psi}>0.4$ at $280\GeV/c$ for
the Feynman-$x$, and a transverse momentum of
$p_{T,\psi\psi}=0.9\pm 0.1\GeV/c$~\cite{Badier:1982ae,Badier:1985ri}. Note that
perturbative QCD neither can explain the NA3 cross section nor the $x_F$
distribution~\cite{Vogt:1995tf} (see also Fig.~\ref{fig:na3data}).

Utilizing the Fermilab negative and positive
charged beams at $600\GeV/c$ to produce charmed particles in a thin foil of
copper or in a diamond, the SELEX collaboration claimed to observe two decay
channels for the $\Xp (ccd)$ state at mass around $3520\MeV/c^2$. The
experiment was operated in the kinematic region $x_F>0.1$. The negative beam
composition was about 50\% $\Sigma^-$ and 50\% $\pi^-$ while the positive
beam was composed of 50\% protons. The experimental data record used both
positive and negative beams. 67\% of the events were induced by $\Sigma^-$,
13\% by $\pi^-$, and 18\% by protons. In the first observation using the
sample of $\Lc\to pK^-\pi^+$~\cite{Kushnirenko:2000ed,Kushnirenko}, a
signal of $15.9$ events over $6.1\pm 0.1$ background events of
$\Xp\to\Lc K^-\pi^+$ was found~\cite{Mattson:2002vu}. To complement this
result, SELEX published an observation of $5.62$ signal events over
$1.38\pm 0.13$ background events for the decay mode $\Xp\to pDK^-$ from a
sample of $D^+\to K^-\pi^+\pi^+$ decays~\cite{Ocherashvili:2004hi}. 

It is interesting to compare the production properties of the $\Xp$ at the
SELEX experiment with the production properties of the double $J/\psi$
production at the NA3 experiment. Unfortunately, it is not possible to
compare the results directly. However, we are still able to compare the
following ratios $R=\sigma(c\bar cc\bar c)/\sigma (c\bar c)$:
\[
R^{\rm SELEX}=R_{\Lc}\times\frac{f(c\to\Lc)}{f_{\Xi_{cc}}}
  \sim(1-4)\times 10^{-3}
\]
and
\[
R^{\rm NA3}=\frac{\sigma(\psi\psi)}{\sigma(\psi)}\times
  \frac{f_{J/\psi}}{f^2_{\psi/\pi}}\sim 2\times 10^{-2},
\]
where $f_{\psi/\pi}\approx 0.03$~\cite{Vogt:1995tf} and
$f_{J/\psi}\approx 0.06$~\cite{Mangano:2004wq}.
$f_{\Xi_{cc}}\approx 0.25$~\cite{Koshkarev:2016rci} represents the fraction
of $cc$ pairs producing the sum of single charged baryons $\Xi_{cc}^+$ and
double charged baryons $\Xi_{cc}^{++}$, but this fraction cannot be less than
the fraction to produce $J/\psi$. Therefore, $R^{\rm SELEX}$ should not be
larger than $10^{-2}$. It is clear that the SELEX production ration is
complemented by the very trustable measurement of the double $J/\psi$
production at the NA3 experiment. It is interesting to note that the intrinsic
charm mechanism predicts $\langle x_F(\Xi_{cc})\rangle=0.33$, as shown in
Ref.~\cite{Koshkarev:2016rci}. This is in excellent agreement with the value
$\langle x_F(\Xp)\rangle\sim 0.33$ measured by the SELEX experiment.

An analysis of the masses of the mass discrepancy between the $\Xp(3520\MeV)$
state measured by SELEX and the $\Xpp(3621\MeV)$ state measured by
LHCb~\cite{Aaij:2017ueg} also complement the SELEX data~\cite{Brodsky:2017ntu}.

\section{Double intrinsic charm at modern experiments}

\subsection{Double $\Jp$ production at the COMPASS experiment}

\begin{figure}[h]
\
\kern-16pt\includegraphics[scale=0.35]{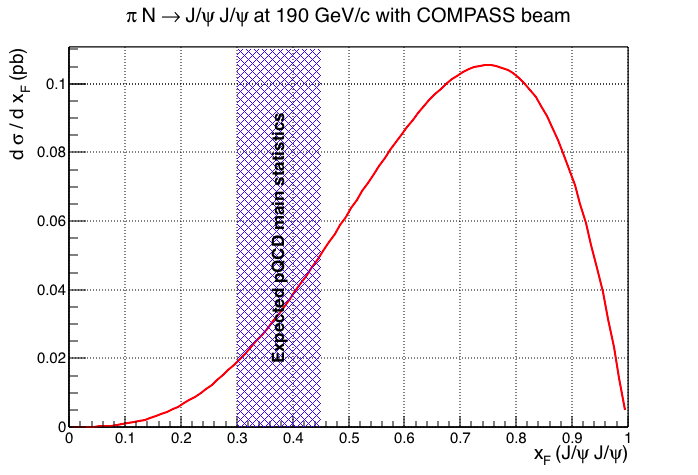}
\caption{\label{fig:compass} Prediction for the double $\Jp$ production at the
COMPASS experiment}
\end{figure}

COMPASS is a fixed-target experiment using the $190\GeV/c$ $\pi^-$ beam at the
CERN SPS. Based on the already collected and planed statistics and compared to
the NA3 statics we can expect up to 100 double $\Jp$
events~\cite{Aghasyan:2017jop} (see also Fig.~\ref{fig:compass}).

\subsection{Production of the doubly charmed baryons at STAR}

The STAR fixed-target program is a fixed-target experiment using the proton
beam of the Relativistic Heavy Ion Collider (RHIC) with up to $250\GeV/c$ and
the Au beam with up to $100\GeV/c$ colliding with a wired target. For
$p_{\rm beam}=200\GeV/c$ we may expect the production cross section of the
$\Xi_{cc}$ to be
$\sigma(\Xi_{cc})\approx(0.2-0.3)\times 75\,\text{nb}$~\cite{Groote:2017szb}.

\begin{figure}%[t]
\kern-16pt\includegraphics[scale=0.35]{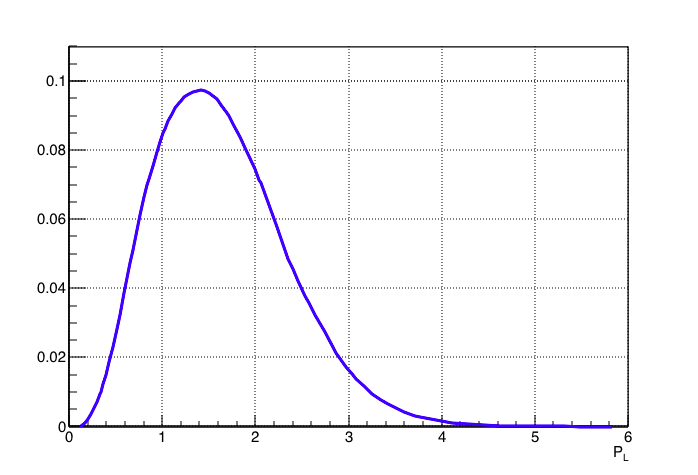}
\kern-16pt\includegraphics[scale=0.35]{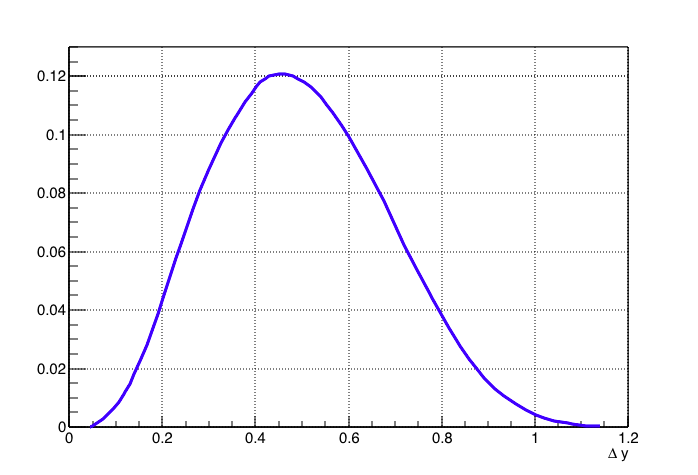}
\caption{\label{fig:xi_cc} Prediction for the $\Xi_{cc}$ momentum distribution
(left panel) and for the distribution of the rapidity difference (right panel)
in the laboratory frame}
\end{figure}

The kinematic limits on the energy and the momentum of the doubly charmed
baryon formed by the intrinsic charm from the target are given by
\begin{equation}\label{eq:kin}
E_{\it lab} = \frac{1}{2m_{\it tar}} (m_{cc}^2 + m_{\it tar}^2),\quad
p_{\it lab} = \frac{1}{2 m_{\it tar}} (m_{cc}^2 - m_{\it tar}^2).
\end{equation}
These last expressions depend solely on the two masses $m_{cc}$ and
$m_{\it tar}$ and no longer on the beam energy. The momentum distribution
and the distribution of the rapidity difference $\Delta y=y-y_{\it tar}$
in the laboratory frame are shown in Fig.~\ref{fig:xi_cc}. It is obvious that
experiments at the STAR detector, typical at rapidities $|y| < 1$ for track
selection, have the potential to observe doubly charmed baryons.

\begin{figure}%[t]
\
\kern-16pt\includegraphics[scale=0.35]{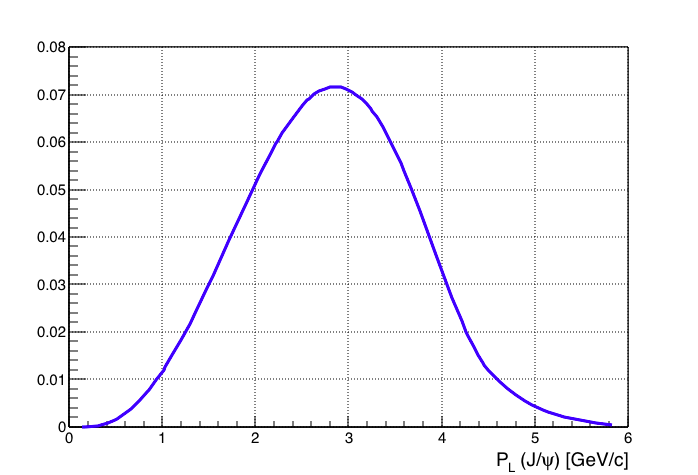}
\kern-16pt\includegraphics[scale=0.35]{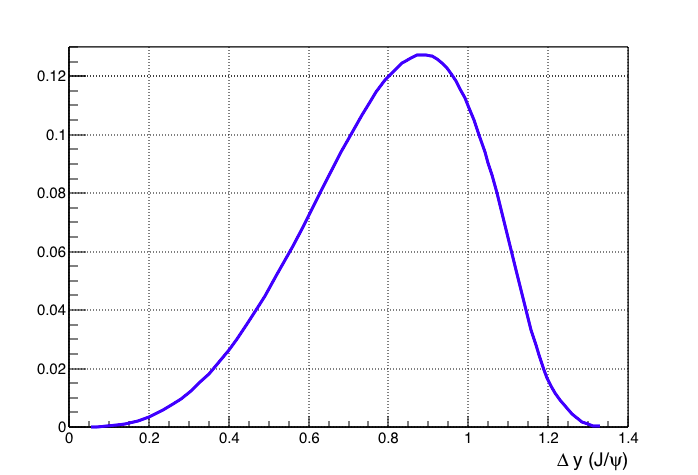}
\caption{\label{fig:jpsi_IC} Prediction for the $\Jp$ momentum distribution
(left panel) and for the distribution of the rapidity difference (right panel)
in the laboratory frame}
\end{figure}

It is interesting as well to present the momentum distribution and the
distribution of the rapidity difference for the $\Jp$ coming from the Fock
state $|uud c\bar c\rangle$ (see Fig.~\ref{fig:jpsi_IC}).

\section{Conclusion}

Current theoretical and experimental knowledge suggests the evidence for the 
existence of intrinsic heavy quarks. Signals of the double intrinsic charm are 
widely accessible at modern experiments. The COMPASS experiment is already
working on the analysis of the double $\Jp$ production~\cite{Guskov}. The STAR
experiment is able to confirm the beautiful prediction of the intrinsic heavy
quark mechanism, producing a heavy quark state nearly at rest.

%%%%

\subsection*{Acknowledgements}
SK would like to thank Alexey Guskov for the invitation at the conference and
much useful advice. This work was supported by the Estonian Research Council
under Grant No.~IUT2-27.

\end{document}